\documentclass[preprint2]{aastex} % for preprints
\usepackage{graphicx}

\shorttitle{Millisecond X-ray Pulsar HETE J1900.1-2455}
\shortauthors{Kaaret et al.}

\begin{document}

\title{Discovery of the Millisecond X-Ray Pulsar HETE J1900.1-2455}

\author{P.\ Kaaret\altaffilmark{1}, E.H.\ Morgan\altaffilmark{2}, R.\
Vanderspek\altaffilmark{2}, J.A.\ Tomsick\altaffilmark{3}}

\altaffiltext{1}{Department of Physics and Astronomy, University of
Iowa,  Van Allen Hall, Iowa City, IA 52242, USA}
%\email{philip-kaaret@uiowa.edu}

\altaffiltext{2}{Center for Space Research, Massachusetts Institute of
Technology, 77 Massachusetts Avenue, Cambridge, MA 02139, USA}

\altaffiltext{3}{Center for Astrophysics and Space Sciences, Code 0424,
9500 Gilman Drive, \\ University of California at San Diego, La Jolla,
CA 92093}

\begin{abstract}

We report the discovery of millisecond pulsations from the low-mass
X-ray binary HETE J1900.1-2455 which was discovered by the detection of
a type I X-ray burst by the High Energy Transient Explorer 2 (HETE-2).
The neutron star emits coherent pulsations at 377.3~Hz and is in an
83.3~minute circular orbit with a companion with a mass greater than
0.016~$M_{\odot}$ and likely less than 0.07~$M_{\odot}$.  The companion
star's Roche lobe could be filled by a brown dwarf with no need for
heating or non-standard evolution.  During one interval with an
unusually high X-ray flux, the source produced quasiperiodic
oscillations with a single peak at $883 \rm \, Hz$ and on subsequent
days, the pulsations were suppressed.  We consider the distribution of
spin versus orbital period in neutron star low-mass X-ray binaries.

\end{abstract}


\section{Introduction}

Shortly after the discovery of radio millisecond pulsars, it was
hypothesized that the neutron stars in low-mass X-ray binaries (LMXBs)
were rotating at hundreds of revolutions per second and are the
progenitors of radio millisecond pulsars \citep{alpar82}.  The presence
of rapidly rotating neutron stars in LMXBs was not conclusively
established until the discovery of coherent millisecond pulsations from
LMXBs \citep{wijnands98,chakrabarty98}. This discovery has greatly
advanced our understanding of these systems.  The unambiguous
measurement of neutron star spin periods was essential in definitely
establishing that the quasiperiodic oscillations found in thermonuclear
X-ray bursts \citep{strohmayer96} are linked to the neutron star spin. 
The highly accurate measurements enabled by the stability of the
neutron star spin have enabled the most accurate measurements of the
orbital properties of LMXBs.  Spectral analysis of the pulse profiles
has the potential to constrain the mass/radius relation of neutron
stars and therefore the equation of state of nuclear matter
\citep{poutanen03}.

However, the sample of millisecond pulsars is small.  Increasing the
sample of millisecond X-ray pulsars is essential to enable population
studies of these systems which should shed light on the properties of
neutron-star X-ray binaries, their evolution, and their relation to
millisecond radio pulsars.  Here, we describe observations made with
the Rossi X-Ray Timing Explorer (RXTE; Bradt, Rothschild, \& Swank
1993) of a new transient X-ray source that we find to be a millisecond
X-ray pulsar: HETE J1900.1-2455 \citep{vanderspek05}.  We describe our
observations in \S 2, our results on the detection of pulsations and
the measurement of the orbital parameters in \S 3, and conclude in \S 4
with a few comments on the nature of the system and the population of
neutron-star LMXBs.

\begin{figure}[tb] \centerline{\epsscale{1.0}\plotone{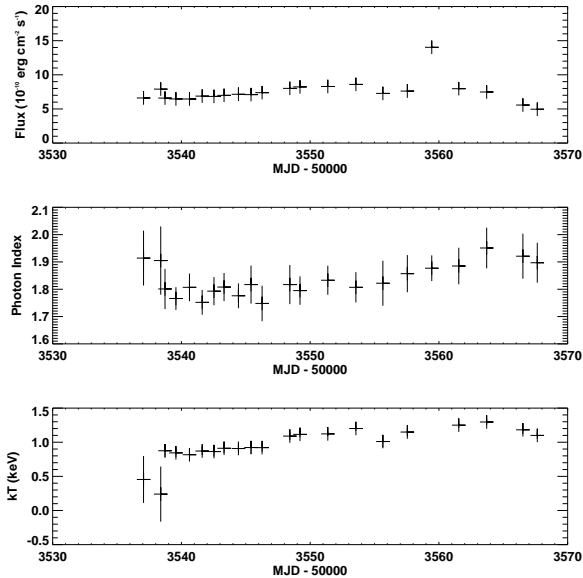}}
\caption{RXTE/PCA light curve of HETE J1900.1-2455.  Each point is the
flux in the 2-20~keV band derived from spectral fitting to an
individual RXTE pointing.} \label{lc} \end{figure}

\begin{figure}[tb] \centerline{\epsscale{1.0}\plotone{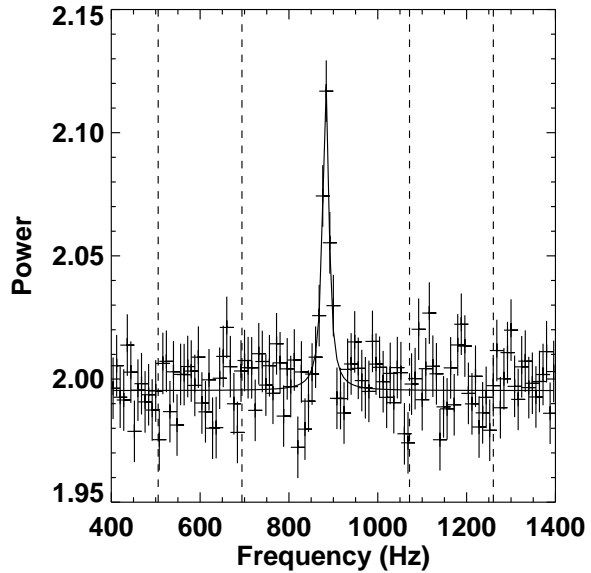}}
\caption{Leahy normalized power spectrum showing detection of a single
kHz QPO.  The solid curve represents the fit of a Lorenztian plus a
constant to the data.  The dashed lines indicate plus and minus the
spin frequency and plus and minus half the spin frequency away from the
kHz QPO centroid frequency.} \label{khzqpo} \end{figure}

\section{Observations of HETE J1900.1-2455}

On 2005 June 14, a bright X-ray burst was detected \citep{vanderspek05}
with the High Energy Transient Explorer 2 (HETE-2) \citep{ricker03}. 
The burst was clearly a type I X-ray burst with radius expansion.  A
detection in the HETE-2 Soft X-Ray Camera (SXC) \citep{villasenor03}
localized the burst to an $80\arcsec$ accuracy.  No known X-ray burst
source was consistent with the SXC position.  Assuming that the burst
luminosity is equal to the  Eddington limit for a $1.4 M_{\sun}$
neutron-star burning helium, \citet{kawai05} estimate the distance as
5~kpc.  

Following discovery of the source, we triggered an RXTE
Target-of-Opportunity observing program which led to multiple
observations on dates from 2005 June 16 to July 16 (MJD 53537 to
53567).  The first few RXTE pointed observations led to the detection
of pulsations at 377.3~Hz \citep{morgan05} and further RXTE
observations enabled an initial measurement of the orbital parameters
\citep{kaaret05}.  A position was determined from RXTE scanning
observations \citep{markwardt05} and then a possible optical
counterpart was identified \citep{fox05}.  The optical counterpart was
confirmed with further optical observations which showed that the
source was bright compared with archival plates, blue in color, and
emits a broad He {\sc ii} emission line \citep{steeghs05}.  An improved
X-ray position, accurate to $5\arcsec$, and consistent with that of the
optical counterpart was obtained with the Swift X-Ray Telescope
\citep{kong05}.  VLA observations were performed, but no radio
counterpart was detected \citep{rupen05}.

For the RXTE observations, we analyzed data from the Proportional
Counter Array (PCA).  The first observation used two single bit modes,
SB\_125us\_8\_13\_1s and SB\_125us\_14\_35\_1s, and an event mode,
E\_16us\_16B\_36\_1s, to provide high resolution timing.  The
subsequent observations used a single event mode, E\_125us\_64M\_0\_1s,
for high resolution timing.  For all observations, the Standard-1
low-resolution (0.125~s) timing mode and the Standard-2 spectroscopic
mode with 16~s time resolution were available.

A light curve derived from the PCA Standard-2 data is shown in
Fig.~\ref{lc}.  We used only data from PCU2 because this PCU was on
during all of the observations and followed the analysis procedures
described in \citet{kaaret02}.  A spectrum was extracted for each
individual observation.  We assign a 1\% systematic uncertainty on each
spectral bin \citep{tomsick99}.  We fit the spectra to a model
consisting of the sum of a power-law and a black body with interstellar
absorption.  We fixed the absorption column density to $N_H = 1.5
\times 10^{21} \rm \, cm^{-2}$ which is the Galactic H{\sc i} column
density along the line of sight.  The source has a Galactic latitude of
$b = -12.87\arcdeg$ and lies 1.1~kpc out of the plane if at a distance
of 5~kpc.  Therefore, the Galactic H{\sc i} should mainly lie between
us and the source.

The discovery X-ray burst occurred on 2005 June 14 (MJD 53535). The
first RXTE pointing occurred 2 days later and shows the source at a
flux of $6.6 \times 10^{-10} \rm \, erg \, cm^{-2} \, s^{-1}$.  The
source flux increases gradually over the beginning of the outburst,
there is one observation with a high flux at MJD 53559, and
subsequently the source flux decays.  The photon index gradually
softens, while the blackbody temperature gradually increases, until a
few days after the peak flux when the trends reverse.  The first two
observations may have had softer and cooler spectra, but the
uncertainties in the spectral fitting do not permit any firm
conclusions in this regard.  We find temperatures higher than those
reported by \citet{campana05}.  We do find that addition of a lower
temperature blackbody improves the fit in some cases.  The true
spectrum of HETE J1900.1-2455 may include two blackbody components, as
has been suggested for other millisecond pulsars \citep{gierlinski05}.

The spectral model described above was inadequate in fitting the high
flux point because that spectrum shows distinct curvature at high
energies.  At energies below 14~keV, the spectrum lies above those from
the previous or subsequent day, while at energies above 14~keV, the
spectrum lies below.  To fit that spectrum, we added a high energy
exponential cutoff and removed the black body component.  The resulting
spectral model is the same as that commonly used for standard X-ray
pulsars, i.e.\ those with spin periods of 1-1000~s
\citep{pravdo78,white83}.  The photon index was $1.88 \pm 0.05$,  the
cutoff energy was $6.43 \pm 0.15 \rm \, keV$, and the folding energy
was $5.87 \pm 0.20 \rm \, keV$.  The cutoff energy is similar to that
found for the X-ray pulsars GS 1843+00 \citep{piraino00} and SMC~X-1
\citep{naik04}, but the photon index is softer than found for those or
other standard accreting X-ray pulsars.  We attempted to fit the
spectrum with a blackbody, the sum of a blackbody plus a power-law, and
the sum of two blackbodies and a power-law, all with absorption, but
were unable to obtain good fits with any of these models.  The RXTE/ASM
light curve does not show unusually high flux near this time.  The
average rate around MJD 53559 is close to 3~c/s, equivalent to roughly
$1.2 \times 10^{-9} \rm \, erg \, cm^{-2} \, s^{-1}$ in the 2-20 keV
band.  It appears unlikely that the high flux point is part of a
superburst.  We examined the light curve of this observation on time
scales of 1, 8 and 64~s.  The light curve shows no usual variability on
these time scales, but does show a kiloHertz quasi-periodic oscillation
(kHz QPO).

We searched for high frequency QPOs in each uninterrupted RXTE
observation window and in combinations of the various data segments. 
We calculated averages of 16~s power spectra including events with
energies below 18~keV from all PCUs.  A single kHz QPO appears in the
observation beginning MJD 53559.447583, see Fig.~\ref{khzqpo}.  The
detection has a significance of $9.2 \sigma$, with no allowance for
trials.  Even allowing for $6 \times 10^{4}$ trials for all of the
distinct frequency ranges in all the power spectra generated, the
detection is still highly significant.  The QPO has a centroid of
$882.8 \pm 1.0 \rm \, Hz$, a width of $16.9 \pm 2.4 \rm \, Hz$, and an
rms amplitude of $0.00787 \pm 0.00086$.  There is no evidence for QPOs
at plus or minus the spin frequency or at plus or minus half the spin
frequency away from the single kHz QPO.

We used the Standard-1 data which has 0.125~s time resolution and no
energy information to search for X-ray bursts.  We found one burst-like
event at MJD 53538.76433.  However, this event appears to be a detector
breakdown event and not an X-ray burst from the source.  HETE-2 has
reported only one additional X-ray burst on the HETE burst summary
page\footnote{http://space.mit.edu/HETE/Bursts/summaries.html} after
the discovery burst.  The additional burst was on MJD 53558, just
before the point in the light curve with a high flux.

We examined the RXTE All-Sky Monitor (ASM) light curve for HETE
J1900.1-2455.  The only detection is of the current outburst. 
Therefore, the recurrence time of the transient is likely longer than
9~years.

\begin{figure}[tb]
\centerline{\epsscale{1.0}\plotone{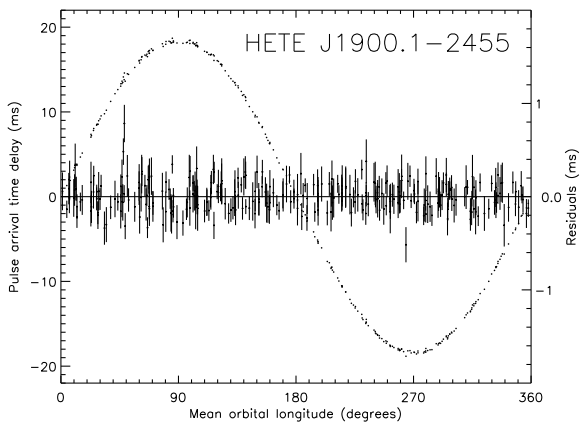}}
%\centerline{\includegraphics[angle=0,width=3.25in]{f3.eps}}
\caption{Pulse timing residuals for HETE J1900.1-2455.  The points
along the sinusoidal curve represent the timing residuals without the
Keplerian orbit correction.  The points centered about zero are the the
timing residuals with the Keplerian orbit correction included and are
multiplied by 10.} \label{orbit} \end{figure}

\begin{deluxetable}{lc}
%\tabletypesize{\scriptsize}
\tablecaption{Parameters of HETE J1900.1-2455 \label{ephem}}
\tablewidth{0pt}
\tablehead{\colhead{Parameter} & \colhead{Value}}
\startdata
Pulse frequency (Hz)                         & 377.296171971(5) \\
Orbital period (s)                           & 4995.258(5) \\
Projected semimajor axis (lt-ms)             & 18.41(1) \\
Epoch of 90$\arcdeg$ mean longitude (MJD TT) & 53549.145385(7) \\
Orbital eccentricity                         & $<$ 0.002 \\
Pulsar mass function ($10^{-6} M_{\sun}$)    & 2.004(3) \\
\enddata
 
\vspace{-12pt}
\tablecomments{These parameters assume the position
reported by \citet{fox05} of $\alpha=$ 19h 00m 08s.65 and $\delta=$
-24$\arcdeg$ 55$\arcmin$ 13$\arcsec$.7 (J2000).  The trailing numbers in
parenthesis indicates the $1-\sigma$ uncertainty in the final digits. 
TT indicates terrestrial time.}
\end{deluxetable}

\section{Pulsations}

We searched for pulsations using the 125~$\mu$s time resolution data by
correcting the event times to the solar system barycenter and then
dividing the data into 256~s intervals, calculating a power spectrum
for events in the 3--8~keV band (channels 8--32 in the 0-255 channel
range) for each interval, and incoherently summing the power spectra
for all intervals within a given pointing.  A signal at 377.3~Hz was
detected in the first pointed RXTE observation of HETE J1900.1-2455,
but not at very high significance.  However, signals near that
frequency were detected in the following four RXTE pointings giving
definite proof of the reality of the pulsations \citep{morgan05}.

We searched the observations between June 17 and 22 for pulsations near
377.3~Hz.  We divided the observations into 256~s intervals and found
the pulse frequency in each interval.  The pulse frequencies had a
clear sinusoidal modulation, characteristic of modulation by orbital
motion.  We fitted a preliminary spin/orbit model to the frequency
measurements to enable a pulse arrival time analysis as is standard in
the timing analysis of pulsars \citep{manchester77}.   In the
preliminary model, we fixed the orbital eccentricity and pulse
frequency derivative to zero.  We corrected the photon arrival times
using the preliminary spin/orbit model, and then calculated the epoch
of the pulse peak within each interval.  We then fitted the pulse
arrival times using a linear least squares regression to find
corrections to the spin and orbital parameters.  The procedure was
iterated as more observations became available and each set of new spin
and orbital parameters were used to recalculate the pulse arrival times
for the full data set.

The orbital and pulse timing parameters presented in Table~\ref{ephem}
were derived using the data from all observations before MJD 53558.  
Fig.~\ref{orbit} shows the residuals of the pulse arrival times with
and without the correction for orbital motion.  A good fit  with
$\chi^2/{\rm DoF} = 1.03$ is obtained without a spin frequency
derivative or orbital eccentricity.  The timing residuals are typically
less than $500\mu$s.  The upper bound on the eccentricity is listed in
the Table.  We were unable to construct a phase-connected timing
solution including data beyond MJD 53558. Observations on MJD 53559
showed the source flux to have brightened dramatically, see above, and
the observed pulse frequency was shifted to 377.291596(16) Hz, $\Delta
\nu/\nu \sim 6 \times 10^{-7}$.  In subsequent observations, pulsations
were suppressed.  The rms pulse amplitude in observations before MJD
53558 ranged from 1.5\% to 5\%, while for observations after MJD 53559,
we place an upper limit on the pulse amplitude of 1\%.

\begin{figure}[tb] \centerline{\epsscale{1.0}\plotone{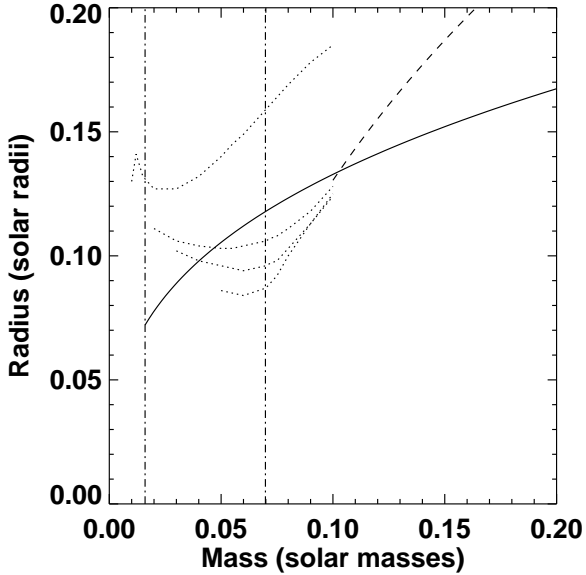}}
\caption{Mass-radius relation for the companion star in HETE
J1900.1-2455.  The solid curve indicates the mass-radius relation for
the companion as required for a Roche-lobe filling companion with the
measured orbital period.  The dashed line represents zero-age main
sequence stars with solar metallicity.  The dotted curves represent
brown dwarfs of ages 0.1 (top), 0.5, 1.0, and 5.0 (bottom)~Gyr.  The
vertical dash-dot lines represent the lower mass limit and the 95\%
confidence upper mass limit.} \label{compan} \end{figure}

\begin{figure}[tb] \centerline{\epsscale{1.0}\plotone{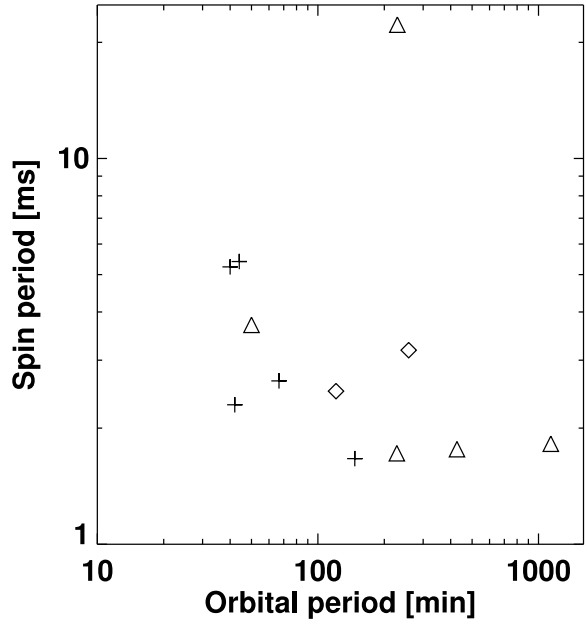}}
\caption{Spin versus orbital period for neutron star X-ray binaries
with spin frequencies greater than 20~Hz -- a `milli-Corbet' diagram. 
Crosses are millisecond pulsars, triangles are burst oscillation
sources, diamonds are sources with both.} \label{mcorbet} \end{figure}

\begin{deluxetable}{lrrll}
%\tabletypesize{\scriptsize}
\tablecaption{Neutron Star Spin Rates and Orbital Periods
\label{nsprop}}
\tablewidth{0pt}
\tablehead{\colhead{Object} & \colhead{Spin Rate} & \colhead{Orbital Period} 
                            & \colhead{Type} & \colhead{Reference}}
\startdata
                 & (Hz) & (min) &    & \\
EXO 0748-676      & 45  &  229  & B  & \citet{villarreal04} \\                
XTE J1807-294     & 191 &  41.1 & P  & \citet{markwardt03a} \\
                  &     &       &    & \citet{markwardt03b} \\
XTE J1751-305     & 435 &  42.4 & P  & \citet{markwardt02} \\
XTE J0929-314     & 185 &  43.6 & P  & \citet{galloway02} \\
4U 1916-05        & 270 &  50.0 & B  &  \\
HETE J1900.1-2455 & 377 &  83.3 & P  & This work \\
SAX J1808.4-3658  & 401 & 120.8 & PB & \citet{wijnands98} \\
                  &     &       &    & \citet{chakrabarty98} \\
IGR J00291+5934   & 599 & 147.4 & P  & \citet{galloway05} \\
4U 1636-536       & 581 &   228 & B  & \\
XTE J1814-338     & 314 & 256.5 & PB & \citet{markwardt03c} \\
X 1658-298        & 567 &   427 & B  & \\
Aql X-1           & 549 &  1137 & B  & \\
\enddata

\vspace{-12pt}
\tablecomments{P indicates millisecond pulsars, B
indicates burst oscillation sources.} 
\end{deluxetable}

\section{Discussion}

Our discovery of coherent millisecond pulsations from HETE J1900.1-2455
adds another to the six previously known X-ray millisecond pulsars
(XMSPs).  As now appears typical for XMSPs, the mass function for the
pulsar is very small, $f_{x} = 2.0 \times 10^{-6} M_{\sun}$, implying
either a very low mass companion or a very improbable orbital
inclination.  The minimum companion mass, assuming a neutron star mass
of $M_X = 1.4 M_{\odot}$ and an orbital inclination $i = 90\arcdeg$, is
$m_2 > 0.016 M_{\odot}$.  The 95\% confidence upper limit on the mass,
assuming a uniform a priori distribution in $\cos i$ and $M_X = 2.2
M_{\odot}$, is $m_2 < 0.07 M_{\odot}$.

If the companion star fills its Roche-lobe, then the density of the
companion star is fixed by the orbital period \citep{frank92}.
Fig.~\ref{compan} shows the mass-radius relation for the companion
star.  Following \citet{bildsten01}, we have plotted the mass-radius
relations for zero-age solar-metallicity main sequence stars
\citep{tout96} and for brown dwarfs of various ages from 0.1 to 5~Gyr
\citep{chabrier00}.  The points at which the stellar mass-radius
relations intersect the mass-radius curve of the companion star
indicate possible stellar companions allowed by the observations.  In
contrast to other XMSPs, there are possible Roche-lobe filling
companion stars with standard evolutionary histories.  Therefore,
heating of the companion star, as suggested for SAX J1808.4-3658
\citep{bildsten01}, is not required to expand the companion's radius
and drive mass transfer in HETE J1900.1-2455.  The companion star is
likely a brown dwarf.

One very unusual aspect of our observations of HETE J1900.1-2455 is the
detection of a dramatic brightening of the source on MJD 53559,
accompanied by a shift in the pulse frequency with $\Delta \nu/\nu \sim
6 \times 10^{-7}$ and then the suppression of pulsations in subsequent
observations.  During the interval of high flux, the X-ray spectrum is
exponentially cutoff at high energies, as is seen for standard X-ray
pulsars, where the accretion spot is thought to be quite compact.  What
mechanism is responsible for the suppression of pulsations after the
high X-ray flux interval is an open question.  Diamagnetic screening of
the magnetic field, a change in the accretion geometry, or enhanced
scattering of the pulses due to accumulation of material around the
poles are possibilities.  Unfortunately, we do not have good
information concerning the total fluence in the high flux event. 
However, the high flux point corresponds to a luminosity of only 1.5\%
of the Eddington luminosity, assuming that the Eddington luminosity is
given by the bolometric flux of $9 \times 10^{-8} \rm \, erg \, cm^{-2}
\, s^{-1}$ reported for the discovery X-ray burst by \citet{kawai05}. 
\citet{cumming01} state that diamagnetic screening of the magnetic
field is not possible for accretion rates below 2\% of the Eddington
rate, which may exclude this possibility.

In addition to the XMSPs, the spin periods of neutron stars in X-ray
binaries can be measured via the detection of (quasiperiodic)
oscillations in type I X-ray bursts.  Table~\ref{nsprop} presents the
spin frequencies and orbital periods for neutron-star low-mass X-ray
binaries with spin rates greater than 20~Hz and measured orbital
periods.  For the millisecond pulsars, the orbital periods are measured
via pulse timing.  For the X-ray bursters, the orbital periods are
determined either via detection of periodic dips in the X-ray emission
or by optical observation of ellipsoidal modulation or Doppler-shifted
absorption lines.  X-ray dips are detected only when the inclination is
high.  The same information is presented in Fig.~\ref{mcorbet}; a
similar plot was presented by \citet{swank04}.  In deference to the
name commonly used for plots of spin versus orbital period for standard
X-ray pulsars \citep{corbet84,corbet86}, we refer to this plot as a
`milli-Corbet' diagram.

There are apparent bounds on the data points at low spin period and at
low orbital period.  The origin of the apparent lower bound on the
orbital period, and the clustering of sources near this limit of
40~min, is a puzzle.  There are LMXBs with shorter orbital periods, but
the neutron star spin frequency has not been definitively measured in
any of those systems.  The lower bound on the spin period distribution
has been interpreted as evidence that gravitational radiation losses
limit the maximum accretion torque spinning up neutron stars in LMXBs
\citep{chakrabarty03}.  The longest spin period, for EXO 0748-676, is a
factor of 4 larger than the second longest.  The unusually slow spin
may be due to an unusually high magnetic field \citep{villarreal04}.

If we consider the neutron stars with spin rates faster than 100~Hz,
there appears to be a lack of points with long spin periods and long
orbital periods.  Excluding EXO 0748-676, the linear correlation
coefficient between the logarithms of the spin and orbital periods is
$r = -0.68$ which has a chance probability of occurrence of 2.5\%. 
This correlation may indicate some coupling between the neutron spin
and orbital period, perhaps indirectly if the mass accretion rate
varies with orbital period.  However, why EXO 0748-676 differs so
significantly from the behavior of the more rapidly rotating neutron
stars would need to be understood.  The discovery of more millisecond
pulsars, or the measurement of the orbital periods of additional X-ray
burst oscillation sources, is needed to better understand the evolution
of neutron star spin in low-mass X-ray binaries.

\acknowledgments  

We greatly appreciate the efforts of the HETE-2 team in the operation
of the HETE-2 satellite and the RXTE team, particularly Evan Smith and
Jean Swank in performing observations and Craig Markwardt in promptly
calculating the clock corrections.  PK thanks Jean Swank for useful
discussions and acknowledges partial support from a NASA grant and a
University of Iowa Faculty Scholar Award.  JAT  acknowledges partial
support from NASA grants NNG04GA49G and NNG04GB19G.

%--------------


\begin{thebibliography}{}

%\bibitem[et al.\ 1981]{616}, ApJ,

\bibitem[Alpar et al.(1982)]{alpar82} Alpar, M.A., Cheng, A.F.,
Ruderman, M. A., \& Shaham, J.\ 1982, Nature, 300, 728

\bibitem[Bildsten \& Chakrabarty(2001)]{bildsten01} Bildsten, L.\ \&
Chakrabarty, D.\ 2001, ApJ, 557, 292

\bibitem[Bradt, Rothschild, \& Swank(1993)]{xte} Bradt, H.V.,
Rothschild, R.E., \& Swank, J.H. 1993, \aaps, 97, 355

\bibitem[Campana et al.(2005)]{campana05} Campana, S.\ et al.\ 2005,
Astron.\ Telegram, No.\ 535

\bibitem[Chabrier et al.(2000)]{chabrier00} Chabrier, G., Baraffe, I.,
Allard, F., \& Hauschildt, P.\ 2000, ApJ, 542, 464

\bibitem[Chakrabarty \& Morgan(1998)]{chakrabarty98} Chakrabarty, D.\
\& Morgan, E.~H. 1998, Nature, 394, 346

\bibitem[Chakrabarty et al.(2003)]{chakrabarty03} Chakrabarty, D.,
Morgan, E.~H., Muno, M.~P., Galloway, D.~K., Wijnands, R., van der
Klis, M., \& Markwardt, C.~B. 2003, Nature, 424, 42

\bibitem[Corbet(1984)]{corbet84} Corbet, R.H.D.\ 1984, A\&A, 141, 91

\bibitem[Corbet(1986)]{corbet86} Corbet, R.H.D.\ 1986, MNRAS, 220, 1047

\bibitem[Cumming, Zweibel, \& Bildsten(2001)]{cumming01} Cumming, A.,
Zweibel, E., \& Bildsten, L.\ 2001, ApJ, 557, 958

\bibitem[Fox(2005)]{fox05} Fox, D.B.\ 2005, Astron.\ Telegram, No.\ 526

\bibitem[Frank, King, \& Raine(1992)]{frank92} Frank, J., King, A., \&
Raine, D.\ 1992, Accretion Power in Astrophysics (Cambridge).

\bibitem[Galloway et al.(2002)]{galloway02} Galloway, D.K.,
Chakrabarty, D., Morgan, E.H., \& Remillard, R.A.\ 2002, ApJ, 576, L137

\bibitem[Galloway et al.(2005)]{galloway05} Galloway, D.K., Markwardt,
C.B., Chakrabarty, D., \& Strohmayer, T.E.\ 2005, ApJ, 622, L45

\bibitem[Gierli\'{n}ski \& Poutanen(2005)]{gierlinski05} 
Gierli\'{n}ski, M.\ \& Poutanen, J.\ 2005, MNRAS, 359, 1261

\bibitem[Kaaret et al.(2002)]{kaaret02} Kaaret, P., in 't Zand, J.J.M.,
Heise, J., \& Tomsick, J.A.\ 2002, ApJ, 575, 1018

\bibitem[Kaaret, Morgan, \& Vanderspek(2005)]{kaaret05} Kaaret, P.,
Morgan, E., \& Vanderspek, R.\ 2005, Astron.\ Telegram, No.\ 538

\bibitem[Kawai \& Suzuki(2005)]{kawai05} Kawai, N.\ \& Suzuki, M.\
2005, Astron.\ Telegram, No.\ 534

\bibitem[Kong, Homan, \& Lewin(2005)]{kong05} Kong, A.K.H., Homan, J.,
\& Lewin, W.H.G.\ 2005, Astron.\ Telegram, No.\ 541

\bibitem[Manchester \& Taylor(1977)]{manchester77} Manchester, R.~N. \&
Taylor, J.~H. 1977, Pulsars (San Francisco: W.~H.~Freeman)

\bibitem[Markwardt et al.(2002)]{markwardt02} Markwardt, C.~B., Swank,
J.~H., Strohmayer, T.~E., in 't Zand, J. J.~M., \& Marshall, F.~E.
2002, ApJL, 575, L21

\bibitem[Markwardt, Smith, \& Swank(2003)]{markwardt03a} Markwardt,
C.~B., Smith, E., \& Swank, J.~H. 2003, IAU Circ., 8080

\bibitem[Markwardt, Juda, \& Swank(2003)]{markwardt03b} Markwardt,
C.~B., Juda, M., \& Swank, J.~H. 2003, IAU Circ., 8095

\bibitem[Markwardt, \& Swank(2003)]{markwardt03c} Markwardt, C.~B., \&
Swank, J.~H. 2003, IAU Circ., 8144, 1

\bibitem[Markwardt et al.(2005)]{markwardt05} Markwardt, C.B, Kaaret,
P, Vanderspek, R., \& Morgan, E.\ 2005, Astron.\ Telegram, No.\ 525

\bibitem[Morgan, Kaaret, \& Vanderspek(2005)]{morgan05} Morgan, E.,
Kaaret, P., \& Vanderspek, R.\ 2005, Astron.\ Telegram, No.\ 523

\bibitem[Naik \& Paul(2004)]{naik04} Naik, S.\ \& Paul, B. 2004, A\&A,
418, 655

\bibitem[Piraino et al.(2000)]{piraino00} Piraino, S.\ et al.\ 2000,
A\&A, 357, 501

\bibitem[Poutanen \& Gierli\'{n}ski (2003)]{poutanen03} Poutanen, J.\
\& Gierli\'{n}ski, M.\ 2003, MNRAS, 343, 1301

\bibitem[Pravdo et al.(1978)]{pravdo78} Pravdo, S.H.\ et al.\ 1978,
ApJ, 225, 988

%\bibitem[Reynolds et al.(1999)]{reynolds99} Reynolds, A.P., Owens, A.,
%Kaper, L., Parmar, A.N., Segreto, A.\ 1999, A\&A, 349, 873

\bibitem[Ricker et al.(2003)]{ricker03} Ricker, G.R.\ et al.\ 2003, in
``Gamma-Ray Burst and Afterglow Astronomy 2001'', eds. G.R.\ Ricker \&
R.\ Vanderspek

\bibitem[Rupen, Mioduszewski, \& Dhawan(2005)]{rupen05} Rupen, M.P.,
Mioduszewski, A.J., \& Dhawan, V.\ 2005, Astron.\ Telegram, No.\ 530

\bibitem[Steeghs et al.(2005)]{steeghs05} Steeghs, D.\ et al.\ 2005,
Astron.\ Telegram, No.\ 543

\bibitem[Strohmayer et al.(1996)]{strohmayer96} Strohmayer, T.E.,
Zhang, W., Swank, J.H., Smale, A., Titarchuk, L., Day, C., \& Lee, U.
1996, \apjl, 469, L9

\bibitem[Swank \& Marshall(2004)]{swank04} Swank, J.H., Marshall, F.E.,
\& the RXTE User's Group, Proposal to the 2004 Senior Review of
Astrophysics Mission Operations \& Data Analysis Programs

\bibitem[Tomsick et al.(1999)]{tomsick99} Tomsick, J.A., Kaaret, P.,
Kroeger, R.A., Remillard, R.A.\ 1999, ApJ, 512, 892

\bibitem[Tout et al.(1996)]{tout96} Tout, C.A., Onno, R.P., Eggleton,
P.P., \& Zhanwen, H.\ 1996, MNRAS, 281, 257

\bibitem[Wijnands \& van der Klis(1998)]{wijnands98} Wijnands, R. \&
van der Klis, M.\ 1998, Nature, 394, 344

\bibitem[Vanderspek et al.(2005)]{vanderspek05} Vanderspek, R., Morgan,
E., Crew, G., Graziana, C., \& Suzuki, M.\ 2005, Astron.\ Telegram,
No.\ 516

\bibitem[Villarreal \& Strohmayer(2004)]{villarreal04} Villarreal,
A.R.\ \& Strohmayer, T.E.\ 2004, ApJ, 614, L121

\bibitem[Villasenor et al.(2003)]{villasenor03} Villasenor, J.N.\ et
al.\ 2003, in ``Gamma-Ray Burst and Afterglow Astronomy 2001'', eds.
G.R.\ Ricker \& R.\ Vanderspek

\bibitem[White, Swank, \& Holt(1983)]{white83} White, N.E, Swank, J.H.,
Holt, S.S.\ 1983, ApJ, 270, 711

\end{thebibliography}
\end{document}